\documentclass[12pt]{iopart}
\usepackage{iopams}
\usepackage{mathbbol}
\usepackage[dvips]{graphicx}

\begin{document}

\title{Quantum spin fluctuations in the dipolar Heisenberg-like rare earth pyrochlores}

\author{Adrian G Del Maestro\dag\ddag and Michel J P Gingras\dag\S}

\address{\dag Department of Physics, University of Waterloo, Waterloo, Ontario, 
              Canada N2L 3G1}
\address{\ddag Department of Physics, Yale University, New Haven, Connecticut,
               USA 06520}
\address{\S Canadian Institute for Advanced Research, Toronto, Ontario, Canada 
            M5G 1Z8}

\ead{adrian.delmaestro@yale.edu}

\begin{abstract}
The magnetic pyrochlore oxide materials of general chemical formula
R$_2$Ti$_2$O$_7$ and R$_2$Sn$_2$O$_7$ (R $\equiv$ rare earth)
display a host of interesting physical behaviours depending on 
the flavour of rare earth ion.  These properties depend on the value
of the total magnetic moment, the crystal field interactions at each rare
earth site and the complex interplay between magnetic exchange and
long-range dipole-dipole interactions.  This work focuses on
the low temperature physics of the dipolar isotropic frustrated antiferromagnetic
pyrochlore materials.  Candidate magnetic ground states are numerically 
determined at zero temperature and the role of quantum spin fluctuations 
around these states are studied using a Holstein-Primakoff spin 
wave expansion to order $1/S$.  The results indicate the strong stability of the 
proposed classical ground states against quantum fluctuations. The inclusion 
of long range dipole interactions causes a restoration of symmetry and a 
suppression of the observed anisotropy gap leading to an increase in quantum 
fluctuations in the ground state when compared to a model with truncated dipole 
interactions. The system retains most of its classical character and there is 
little deviation from the fully ordered moment at zero temperature.
\end{abstract}

\submitto{\JPCM}
\maketitle


\newcommand{\Gd}{\mathrm{Gd}_2\mathrm{Ti}_2\mathrm{0}_7}
\newcommand{\Tb}{\mathrm{Tb}_2\mathrm{Ti}_2\mathrm{0}_7}

\newcommand{\tha}{\theta_\alpha}
\newcommand{\thb}{\theta_\beta}
\newcommand{\Cm}[1]{\mathrm{C}^{(#1)}_{\alpha \beta}}
\newcommand{\ea}[1]{\eta_{\alpha}^{#1}}
\newcommand{\eb}[1]{\eta_{\beta}^{#1}}
\newcommand{\er}{\rme^{i\bi{k}(\ra - \rb)}}
\newcommand{\edot}[3]{\exp[{{#1}\rmi{#2}\cdot{#3}}]}
\newcommand{\vd}[2]{\bi{#1}_{#2}}
\newcommand{\vu}[2]{\bi{#1}^{#2}}
\newcommand{\mc}[1]{\mathcal{#1}}
\newcommand{\ek}[1]{\varepsilon_\alpha ({#1}\vk)}
\newcommand{\et}[2]{\exp[ {#1} \ek{#2} \diagup T] }

\newcommand{\lbr}{\left (}
\newcommand{\rbr}{\right )}
\newcommand{\lf}{\left}
\newcommand{\rt}{\right}

\newcommand{\Svra}{\bi{S}_{\alpha}(\bi{R}^{\mu})}
\newcommand{\Svrat}{\tilde{\bi{S}}_{\alpha}(\bi{R}^{\mu})}
\newcommand{\Svrb}{\bi{S}_{\beta}(\bi{R}^{\nu})}
\newcommand{\Smr}{\mathrm{S}}
\newcommand{\Sc}[2]{S^{#1}_{#2}}
\newcommand{\Sket}[1]{\ket{\Smr,s^z_{#1}}}

\newcommand{\Scra}[1]{S^{#1}_{\alpha}(\bi{R}^{\mu})}
\newcommand{\Scrb}[1]{S^{#1}_{\beta}(\bi{R}^{\nu})}
\newcommand{\Stcra}[1]{\tilde{S}^{#1}_{\alpha}(\bi{R}^{\mu})}
\newcommand{\Stcrb}[1]{\tilde{S}^{#1}_{\beta}(\bi{R}^{\nu})}
\newcommand{\Stcka}[1]{\tilde{S}^{#1}_{\alpha}(\bi{k})}
\newcommand{\Stckb}[2]{\tilde{S}^{#2}_{\beta}({#1}\bi{k})}
\newcommand{\Stcoa}[1]{\tilde{S}^{#1}_{\alpha}(0)}
\newcommand{\Stckbp}[1]{\tilde{S}^{#1}_{\beta}(\bi{k}')}

\newcommand{\ra}{\bi{r}_\alpha}
\newcommand{\rb}{\bi{r}_\beta}
\newcommand{\rab}{\bi{r}_{\alpha \beta}}
\newcommand{\rn}[1]{\bi{r}^{#1}_{\alpha \beta}}
\newcommand{\R}{\bi{R}_{\alpha \beta}^{\mu \nu}}
\newcommand{\Rc}{\boldsymbol{\mathcal{R}}_{\alpha \beta}^{n}}
\newcommand{\nrm}[2]{\left |{#1} \right |^{#2}}
\newcommand{\nhga}[1]{\bi{n}_{{#1},\alpha}}
\newcommand{\nhia}{\bi{n}_{i,\alpha}}
\newcommand{\nhjb}{\bi{n}_{j,\beta}}
\newcommand{\ncia}[1]{n^{#1}_{i,\alpha}}
\newcommand{\ncjb}[1]{n^{#1}_{j,\beta}}
\newcommand{\nndot}{\lf(\nhia \cdot \nhjb \rt)}
\newcommand{\nrdot}{\lf(\nhia \cdot \R \rt)\lf(\nhjb \cdot \R \rt)}
\newcommand{\ndotda}{\lf(\nhia \cdot \bi{\nabla}_{\bi{x}} \rt)}
\newcommand{\ndotdb}{\lf(\nhjb \cdot \bi{\nabla}_{\bi{x}} \rt)}
\newcommand{\vx}{\bi{x}}
\newcommand{\vk}{\bi{k}}
\newcommand{\vg}{\bi{G}}

\newcommand{\px}[1]{\cfrac[l]{\partial}{\partial x} \genfrac{}{}{0pt}{}{}{^#1}}
\newcommand{\barx}{\rt|_{\bi{x}=0}}

\newcommand{\dlta}[2]{\delta_{{#1},{#2}}}

\newcommand{\Jce}[3]{J^{{#1} {#2}}_{\alpha \beta}\left({#3}\right)}
\newcommand{\Jc}[3]{\mc{J}^{{#1} {#2}}_{\alpha \beta}\left({#3}\right)}
\newcommand{\gub}{g\mu_{\mathrm{B}}}
\newcommand{\J}[1]{\mathcal{J}_{\alpha \beta}^{i j}\left({#1}\right)}
\newcommand{\D}[1]{\mathcal{D}_{\alpha \beta}^{i j}\left({#1}\right)}
\newcommand{\E}[1]{\mathcal{E}_{\alpha \beta}^{i j}\left({#1}\right)}
\newcommand{\K}[1]{\mathcal{K}_{\alpha \beta}^{i j}\left({#1}\right)}
\newcommand{\W}{\mathcal{W}_{\alpha \beta}^{i j}\left({\vk}\right)}
\newcommand{\X}{\mathcal{X}_{\alpha \beta}^{i j}\left({\vk}\right)}
\newcommand{\Y}{\mathcal{Y}_{\alpha \beta}^{i j}\left({\vk}\right)}
\newcommand{\U}[2]{\mathcal{U}{#1}_{\alpha \beta}^{i j}\left({#2}\right)}
\newcommand{\Ddd}{D_{dd}}

\newcommand{\smk}{\sum_{\bi{k}}}
\newcommand{\smab}{\sum_{\alpha, \beta}}
\newcommand{\smij}{\sum_{i, j}}
\newcommand{\smuv}{\sum_{\mu, \nu}}
\newcommand{\sms}{\smab\smij\smuv}
\newcommand{\smsk}{\smab\smij\smk}
\newcommand{\sma}{\sum_\alpha}

\newcommand{\ca}{c_\alpha (\bi{R}^\mu)}
\newcommand{\cda}{c^\dagger_\alpha (\bi{R}^\mu)}
\newcommand{\cb}{c_\beta (\bi{R})^\nu}
\newcommand{\cdb}{c^\dagger_\beta (\bi{R}^\nu)}

\newcommand{\cak}[1]{c_\alpha ({#1}\bi{k})}
\newcommand{\cdak}[1]{c^\dagger_\alpha ({#1}\bi{k})}
\newcommand{\ak}[1]{a_\alpha ({#1}\bi{k})}
\newcommand{\adk}[1]{a^\dagger_\alpha ({#1}\bi{k})}
\newcommand{\cbk}[1]{c_\beta ({#1}\bi{k})}
\newcommand{\cdbk}[1]{c^\dagger_\beta ({#1}\bi{k})}
\newcommand{\aks}[2]{a_{#2} ({#1}\bi{k})}
\newcommand{\adks}[2]{a^\dagger_{#2}({#1}\bi{k})}

\newcommand{\Qc}[2]{\mathsf{Q}_{{#1},{#2}}}
\newcommand{\Qdc}[2]{\mathsf{Q}^{\dagger}_{{#1},{#2}}}

\newcommand{\csa}{\cos\tha}
\newcommand{\csb}{\cos\thb}
\newcommand{\sna}{\sin\tha}
\newcommand{\snb}{\sin\thb}
\newcommand{\cosp}{\cos\frac{\pi}{2}(\alpha + \beta)}
\newcommand{\cosm}{\cos\frac{\pi}{2}(\alpha - \beta)}
\newcommand{\cosn}[1]{\cos q_{#1}}
\newcommand{\cosnm}[2]{\cos (q_{#1} - q_{#2})}

\newcommand{\Gma}[2]{\Gamma^{{#1}{#2}}_{\alpha \beta}}

\newcommand{\Nd}[1]{\bi{\mathsf{#1}}^{\dagger}}
\newcommand{\N}[1]{\bi{\mathsf{#1}}}
\newcommand{\Nab}[3]{\mathsf{#2}_{\alpha \beta}^{#1}({#3})}

\section{Introduction}
\label{intro}
Magnetic frustration arises when magnetic moments (spins) are unable
to minimize their classical ground state energy by minimizing the two body 
magnetic exchange interactions pair by pair. The simplest example of
magnetic frustration is for antiferromagnetically coupled spins on a
triangular lattice where the triangular coordination prevents the spins
from pointing mutually antiparallel to each other and where,
consequentially, the spins adopt a non-collinear structure to minimize
their energy.  Due to the canted structure, the internal mean-field
is much smaller in frustrated spin systems than in conventional
collinear magnets. As a result, frustrated magnetic systems have an
enhanced propensity for large zero temperature quantum spin fluctuations
allowing for the possibility of exotic quantum ground states 
\cite{Chandra, Can-Lac}.

The pyrochlore lattice of corner-sharing tetrahedra where spins
interact via nearest neighbour antiferromagnetic exchange interactions
has attracted much interest over the past fifteen years \cite{pyro}.
In particular, the R$_2$Ti$_2$O$_7$ and
R$_2$Sn$_2$O$_7$ magnetic insulators (where R is a trivalent rare-earth
ion, R=Tb, Ho, Dy, Gd, Yb and Er, sitting on a regular pyrochlore
lattice) have been found to display a multitude of magnetic phenomena.
For example, the combination of the strong Ising anisotropy in
Dy$_2$Ti$_2$O$_7$ \cite{ram-nat,MJP-BDH}, Ho$_2$Ti$_2$O$_7$ 
\cite{harris,bramwell}, and Ho$_2$Sn$_2$O$_7$ \cite{HoSnO} gives rise to 
``spin ice'' behavior where the system exhibits
extensive magnetic entropy and maps onto the problem of proton disorder
in common ice water \cite{spin-ice1,spin-ice2}. In Yb$_2$Ti$_2$O$_7$, a
sharp first order transition is observed in specific heat measurements,
but with no obvious sign of long range order in neutron scattering
\cite{YbTiO}.  The Tb$_2$Ti$_2$O$_7$ antiferromagnet is particularly
interesting since it appears to be an excellent candidate for
collective paramagnetism \cite{villain} or ``spin liquid'' behavior,
as it fails to develop long range order down to 50~mK despite a
Curie-Weiss temperature of $\theta_{\rm CW}= -20$~K
\cite{TbTiO-1,TbTiO-2,TbTiO-3}.  Finally, both Er$_2$Ti$_2$O$_7$ \cite{ErTiO} and
Gd$_2$Ti$_2$O$_7$ \cite{gdtio-raju,gdtio-huse,gdtio-bram} display a transition to 
long-range ordered states below 1~K.

In the Ho$_2$Ti$_2$O$_7$, Dy$_2$Ti$_2$O$_7$, Tb$_2$Ti$_2$O$_7$ and
Gd$_2$Ti$_2$O$_7$ materials, the rare-earth ion carries a large magnetic
moment between five and ten Bohr magnetons and the dipolar energy scale at
nearest neighbour makes a sizeable contribution to the total magnetic
interaction. Indeed, in Dy$_2$Ti$_2$O$_7$ and
Ho$_2$Ti$_2$O$_7$, it has been shown that it is the dipolar
interactions that cause the spin ice phenomenology \cite{MJP-BDH,spin-ice1}. In
Tb$_2$Ti$_2$O$_7$, the dipolar interactions play a very important role
in frustrating the nearest neighbour exchange interactions and are a
significant player in the formation of a possible spin liquid state \cite{TbTiO-3}. 
In Gd$_2$Ti$_2$O$_7$, the $4f$ orbital of Gd$^{3+}$ is
half filled, and crystal-field single ion anisotropy effects are
negligible.  This system, with $\theta_{\rm CW}\sim $ -9.8~K,  and spin
quantum number $S=7/2$ therefore constitutes a useful starting point as
a real material which could be described by a nearest neighbour Heisenberg 
pyrochlore model \cite{Can-Lac,Moes-Chalk}.  However, in this system 
the large dipolar interactions are important since they have been predicted 
to select a unique long-range ordered classical ground state \cite{gdtio-huse,Palm-Chalk}  

Rare-earth pyrochlores do not hold a monopoly on unusual phenomena among 
the insulating rare-earth magnetic materials. Possibly one of the most 
interesting cases is that of the Gd$_3$Ga$_5$O$_{12}$ three dimensional garnet 
of corner-sharing triangles \cite{GGG1}. In this material, experimental 
evidence of a magnetic field induced spin liquid has been found \cite{GGG1}, as well 
as reentrance from a long range ordered magnetic state to a spin 
liquid upon cooling \cite{GGG2,GGG3}.

The discussion in the previous two paragraphs brings up the following
question. In conventional exchange-coupled Heisenberg antiferromagnets
the quantum fluctuations, $\Delta S$, are governed by an integral over
the Brillouin zone of $1/\omega(\bi{k})$ where $\omega(\bi{k})$ are the
$\bi{k}$-dependent excitation frequencies.  Weakly dispersive modes give 
rise to larger fluctuations and (anisotropy) gapped modes contribute smaller 
quantum fluctuations \cite{kittel}.  In the nearest neighbour Heisenberg pyrochlore,
we have two zero energy modes throughout the zone and any conventional
perturbative spin wave calculation is therefore singular. In systems
such as the rare-earth pyrochlores above, we expect both single ion
anisotropy and dipolar interactions to lift the classical degeneracy
and make the spin excitations dispersive. The leading effect of those
perturbations will be to push the two zero modes throughout the zone to
finite frequency. This is akin to the finite energy $\bi{k}$-independent
resonance observed in inelastic neutron scattering in the ZnCr$_2$O$_4$
spinel \cite{ZnCrO}.  In other words, in the isotropic
Heisenberg antiferromagnet with conventional long range N\'eel order,
it is the dispersion of the acoustic magnons ($\omega(\bi{k}) \sim \nrm{\bi{k}}{}$)
that controls the leading $1/S$ quantum fluctuations. The question therefore concerns
the magnitude of quantum fluctuations arising from gapped excitations with negligible
dispersion throughout the Brillouin zone in a highly frustrated system of the type 
considered here.  This would seem particularly relevant to Gd$_2$Ti$_2$O$_7$ where the
experimentally observed ground state seems to be at odds with classical predictions
\cite{gdtio-bram,Palm-Chalk}.

The outline of the paper is as follows. In Section~2 we discuss the
general framework for calculating the quantum spin fluctuations in a
non-Bravais lattice in the presence of combined exchange and long range
dipolar interactions. In Section~3, we apply the method discussed in
Section~2 to the specific case of the Gd$_2$Ti$_2$O$_7$ pyrochlore
material.  Section~4 contains a brief conclusion of our work.

\section{$1/S$ Formalism}
\label{formalism}
In this section we introduce the Hamiltonian believed to describe our antiferromagnetic
pyrochlore system and map out its low temperature excitations using linear spin wave theory. 
From the diagonalized form of the spin wave Hamiltonian we obtain expressions for the zero
point energy shift and reduction in the classical staggered moment due to spin fluctuations.
Utilizing the partition function of the non-interacting Bose gas, the relevant thermodynamic
quantities are given in terms of the calculated spin wave dispersion energies.

\subsection{Model Hamiltonian}
A general two-body spin interaction Hamiltonian including 
both p$^{\rm th}$ nearest neighbour anisotropic exchange and long range 
dipole-dipole interactions can be written as
\begin{eqnarray}
\fl \mc{H} = -\frac{1}{2}\sum_{\mu,\alpha}\sum_{\nu,\beta}\sum_{k,l}
           \Jce{k}{l}{\R}\Scra{k}\Scrb{l} \nonumber \\
           \lo{+} \frac{1}{2}\Ddd\sum_{\mu,\alpha}\sum_{\nu,\beta} \lf[
	    \frac{\Svra\cdot\Svrb}{\nrm{\R}{3}} - 3\frac{\lf(\Svra\cdot\R\rt)
	    \lf(\Svrb\cdot\R\rt)}{\nrm{\R}{5}}\rt]
\label{full-ham}
\end{eqnarray}
using $\R = (\vu{R}{\nu}+\rb) - (\vu{R}{\mu}+\ra)$, $\Jce{k}{l}{\R} = 
\sum_{n=1}^{p} J^{k l}_n f^n_{\alpha \beta} \dlta{\R}{\Rc}$,
($\Rc$ is the set of n$^{\rm th}$ nearest neighbour connection vectors) and 
$\Ddd = \mu_0(\gub)^2/4\pi$ where the spins are full O(3) operators satisfying 
$\Svra\cdot\Svra = \Smr(\Smr+1)$.  A factor of $\frac{1}{2}$ has been included to 
avoid double counting.  The anisotropic exchange tensor $J^{kl}_n$ is the relative 
strength of the magnetic interaction for spin components $k$ and $l$ between 
n$^{\rm th}$ nearest neighbours.  The factor $f^n_{\alpha,\beta}$ is equal to 
$1-\dlta{\alpha}{\beta}$ if the coupled spins are on
different sublattices, or $\dlta{\alpha}{\beta}$ if they are on the same sublattice.
The symbols $\mu$ and $\nu$ index all primitive Bravais lattice vectors, $\alpha$ and $\beta$
run over all basis vectors and $k$ and $l$ over the spin components $x, y$ and $z$. We wish
to consider the stability of some classical ground state by investigating the role of
quantum fluctuations in reducing the fully polarized classical spin value $\bi{S} =
(0,0,\Smr)$.  This can be accomplished by changing the axis of quantization from the
global $z$-direction (which is an arbitrary choice of the theory) to a local quantization 
axis which points in the direction of each spin.  Let $\tilde{\bi{S}}_{\alpha}(\vu{R}{\mu})$ 
be the locally quantized spin which is related to the real spin operator defined in the 
crystallographic frame via the rotation
\begin{equation}
\Svra = \N{R}(\theta_\alpha, \phi_\alpha)\tilde{\bi{S}}_{\alpha}(\vu{R}{\mu}).
\label{spin-rotation}
\end{equation}
This transformation can be implemented in an alternate way by defining an orthogonal triad
of unit vectors $\nhia$ for each sublattice $\alpha$ by
\begin{equation}
\nhia = \N{R}(\theta_\alpha, \phi_\alpha)\bi{e}_i
\end{equation}
where $\{\bi{e}_x,\bi{e}_y,\bi{e}_z\}$ are the usual Cartesian unit vectors and 
clearly $\bi{n}_{z,\alpha}$ points in the direction of the spin $\Svra$.  In this 
form, we can re-express \Eref{spin-rotation} as
\begin{equation}
\Svra = \sum_{i}\Stcra{i}\nhia
\end{equation}
and substituting into our original Hamiltonian \Eref{full-ham} we find,
\begin{equation}
\mc{H} = -\frac{1}{2}\sms \J{\R}\Stcra{i}\Stcrb{j}
\label{red-ham}
\end{equation}
where
\begin{equation}
\fl\J{\R} = \sum_{k,l}\Jce{k}{l}{\R}\ncia{k}\ncjb{l}
       -\!\Ddd \lf[ \frac{\nndot}{\nrm{\R}{3}} - 3\frac{\nrdot}{\nrm{\R}{5}}\rt]
\label{RS-J}
\end{equation}
is the two spin coupling matrix.  Making use of the periodicity of the lattice 
we Fourier transform the spin operators via
\begin{equation}
\Stcra{i} = \frac{1}{\sqrt{N}}\smk \Stcka{i}\edot{}{\bi{k}}{(\vu{R}{\mu}+\ra)},
\label{spin-FT}
\end{equation}
where $N$ is the number of Bravais lattice points, leading to
\begin{eqnarray}
\fl\mc{H} = -\frac{1}{2}\sms \sum_{\bi{k},\bi{k}'} \J{\R}\Stcka{i}\Stckbp{j} \nonumber \\
	    \times \edot{}{\bi{k}}{(\vu{R}{\mu}+\ra)}\edot{}{\bi{k}'}{(\vu{R}{\nu}+\rb)}
\end{eqnarray}
and using the definition of the Dirac delta function $\dlta{\bi{q}}{\bi{q}'} =
\frac{1}{N}\sum_{\upsilon}\edot{}{(\bi{q}-\bi{q}')}{\vu{R}{\upsilon}}$,
we can write this as
\begin{equation}
\mc{H} = -\frac{1}{2}\smsk\Stcka{i}\J{\bi{k}}\Stckb{-}{j}
\label{k-space-ham}
\end{equation}
where the Fourier transform of the spin-spin interaction matrix has been defined by
\begin{equation}
\J{\bi{k}} = \sum_{\mu}\J{\R}\edot{-}{\bi{k}}{\R}.
\end{equation}


The Holstein-Primakoff transformation \cite{holstprim} defined by 
\numparts
\begin{eqnarray}
\Stcra{+} = \sqrt{2\Smr - \cda\ca}\ \ca  \\
\Stcra{-} = \cda\ \sqrt{2\Smr - \cda\ca} \\
\Stcra{z} = \Smr - \cda\ca
\label{spm-exp}
\end{eqnarray}
\endnumparts
allows us to express our anticommuting fermionic spin operators in terms of commuting
bosonic spin deviation annihilation and creation operators $\ca$ and $\cda$ where 
$[\ca,c^\dag_\beta(\vu{R}{\nu})] = \dlta{\alpha}{\beta}\dlta{\mu}{\nu}$.  We wish to 
deal with a theory whose transverse terms are only linear in the boson operators, leading
to a quadratic boson-mode spin wave Hamiltonian.  In other words, a harmonic theory where there
exists only small deviations away from some stable minimum energy spin configuration. 
Identifying that the boson operators can be written in terms of their Fourier components
\numparts
\begin{eqnarray}
\ca   = \frac{1}{\sqrt{N}} \smk \cak{} \edot{-}{\vk}{(\vu{R}{\mu}+\ra)} \\
\cda  = \frac{1}{\sqrt{N}} \smk \cdak{} \edot{}{\vk}{(\vu{R}{\mu}+\ra)},
\label{boson-FT}
\end{eqnarray}
\endnumparts
we can write the linearized Holstein-Primakoff transformation in reciprocal space as
\numparts
\begin{eqnarray}
\Stcka{x} & = \sqrt{\frac{\Smr}{2}}\lf[\cdak{} + \cak{-} \rt] + \Or\lf(\frac{1}{\Smr^2}\rt) \\
\Stcka{y} & = \rmi\sqrt{\frac{\Smr}{2}}\lf[\cdak{} - \cak{-} \rt] + \Or\lf(\frac{1}{\Smr^2}\rt) \\
\Stcka{z} & = \sqrt{N}\Smr\dlta{\vk}{0}\ \edot{-}{\vk}{\ra} - 
	      \frac{1}{\sqrt{N}}\sum_{\vk'}c^\dagger_\alpha(\vk') c_\alpha(\vk'-\vk),
\label{HP-FT}
\end{eqnarray}
\endnumparts
where the Fourier transformed boson operators satisfy $[\cak{},\ c^\dagger_\beta(\bi{k}')]
=\dlta{\alpha}{\beta}\dlta{\bi{k}}{\bi{k}'}$.

Performing this transformation on \Eref{k-space-ham} and keeping only terms up to second
order in the boson mode amplitudes, we obtain a form for the Hamiltonian to order
$1/S$ which can be broken into three parts that are zeroth, first and second order 
in the boson operators.
\begin{equation}
\mc{H} = \mc{H}^{(0)} + \mc{H}^{(1)} + \mc{H}^{(2)}
\label{H}
\end{equation}
with
\numparts
\begin{eqnarray}
\fl\mc{H}^{(0)}  = -\frac{1}{2}N\Smr(\Smr+1)\smab\Jc{z}{z}{0} \label{H-0} \\
\fl\mc{H}^{(1)}  = -\Smr\sqrt{\frac{N\Smr}{2}} \smab \lf[\Nab{}{F}{0}c^\dagger_\alpha(0) +
                   \Nab{\ast}{F}{0}c_\alpha(0) \rt] \label{H-1}\\
\fl\mc{H}^{(2)}  = -\frac{1}{2}\Smr\smab\smk \lf[\Nab{}{A}{\bi{k}}\cdak{}\cbk{} +
		  \Nab{}{B}{\bi{k}}\cdak{}\cdbk{-} \rt. \nonumber \\
	     \lf. +\ \Nab{\ast}{B}{-\bi{k}}\cak{-}\cbk{}+\Nab{\ast}{A}{-\bi{k}}\cak{-}\cdbk{-} \rt]
\label{H-2}
\end{eqnarray}
\endnumparts
and
\numparts
\begin{eqnarray}
\fl\Nab{}{F}{0} & = \Jc{x}{z}{0} + \rmi\Jc{y}{z}{0} \\
\fl\Nab{}{A}{\bi{k}} & = \frac{1}{2} \lf\{ \Jc{x}{x}{\bi{k}} + \Jc{y}{y}{\bi{k}} - 
                       \rmi\lf[ \Jc{x}{y}{\bi{k}} - \Jc{y}{x}{\bi{k}} \rt] \rt\}
		      -\sum_\gamma \mc{J}^{zz}_{\alpha\gamma}(0) \dlta{\alpha}{\beta}
		      \label{A} \\
\fl\Nab{}{B}{\bi{k}} & = \frac{1}{2} \lf\{ \Jc{x}{x}{\bi{k}} - \Jc{y}{y}{\bi{k}} + 
                       \rmi\lf[ \Jc{x}{y}{\bi{k}} + \Jc{y}{x}{\bi{k}} \rt] \rt\}.
\label{B}
\end{eqnarray}
\endnumparts
Here we note the manifestation of the equilibrium condition, that for a particular 
classical ground state to be stable, the effective magnetic field acting on each spin is 
parallel to that spin.  This implies that there are no terms linear in the boson operators
in \Eref{H}, or $\sum_{\beta}\Nab{}{F}{0} = 0$ showing that the 
$\Jc{x}{z}{0}, \Jc{y}{z}{0}, \Jc{z}{x}{0}\mbox{ and }
\Jc{z}{y}{0}$ blocks of the interaction matrix are only responsible for determining the
equilibrium configuration and do not play any dynamical role \cite{magnetism-I}.


\subsection{Diagonalization}
If a stable ground state spin-configuration exists, then the quadratic spin wave Hamiltonian 
in \Eref{H-2} can be rewritten in terms of a general quadratic form by defining a 
vector which contains the boson creation and destruction operators
\begin{equation}
\N{X} = \lf[c_1 (\vk) \ldots c_s(\vk),\ c_1^\dagger (-\vk) \ldots c_s^\dagger (-\vk)
	    \rt]^{T}
\end{equation}
where $s$ is the number of atoms in the basis. In matrix notation, this leads to 
the Hamiltonian
\begin{equation}
\mc{H} = \mc{H}^{(0)} + \smk \Nd{X} \N{H} \N{X}
\label{H-mat}
\end{equation}
where $\N{H}$ is the $2s\times2s$ matrix defined by
\begin{equation}
\N{H} = -\frac{1}{2}\Smr 
\lf[\begin{array}{cc}
\mathsf{A} (\vk)         & \mathsf{B} (\vk) \\
\mathsf{B}^{\ast} (-\vk) & \mathsf{A}^{\ast} (-\vk) 
\end{array}\rt]
\end{equation}
and $\mathsf{A} (\vk)$ and $\mathsf{B} (\vk)$ are the full matrix representations of
\Eref{A} and (\ref{B}).  We now introduce the transformation 
$\N{X} = \N{Q}\N{Y}$ where $\N{Y}$ is a column matrix of new boson operators 
$a_\alpha (\vk)$ and $a^\dag_\alpha (\vk)$.  It is clear that if we 
intend to diagonalize our spin wave Hamiltonian, the transformation law
must satisfy $\Nd{Q} \N{H} \N{Q} = \bLambda$ where $\bLambda$ is the diagonal matrix
of eigenenergies. The matrix $\N{Q}$ constructed from the eigenvectors of the 
transformation must be normalized with respect to a metric which preserves proper 
boson commutation rules for our new operators.  This requirement leads to the 
possibility of a non-unitary $\N{Q}$, and in general forces us to perform a Bogoliubov 
transformation to diagonalize the spin wave Hamiltonian \cite{turov-ppmoc}.  The details of 
the numerical construction of this transformation are included in \ref{ncbt} and the 
result is given by
\begin{equation}
\fl\mc{H} = \mc{H}^{(0)} + \smk \sum_\alpha \ek{} + \smk \sum_\alpha \ek{} 
	    \lf[\adk{} \ak{} + \adk{-} \ak{-} \rt].
\label{QBSWH}
\end{equation}


\subsection{Zero Point Effects}
In a quantum mechanical antiferromagnet, there exists spin fluctuations
at zero temperature with energy $\varepsilon(\bi{k})$ which raise the classical
ground state energy and reduce the staggered magnetic moment per spin from 
its classical value of $\Smr$. Examining \Eref{QBSWH} and noting the fact that
at zero temperature, $\lf\langle \adk{} \ak{} \rt\rangle = 0$, the contribution to the 
ground state energy from quantum spin fluctuations to order $1/\Smr$ will be given by
\begin{equation}
\Delta \mc{H}^{(0)} = \smk \sum_\alpha \ek{}.
\label{DeltaH}
\end{equation}

Returning to the original Holstein-Primakoff bosons, the reduction in the classical spin 
polarization will be equivalent to the average number of activated magnons 
localized at each site,  
\begin{equation}
\Delta \Smr = \frac{1}{Ns} \sum_\mu \sum_\alpha \left \langle \cda\ca \right \rangle
\end{equation}
where $\langle \cdots \rangle$ indicates the expectation value with respect to the
classical ground state.  Upon moving to Fourier space, this becomes an average over
the first Brillouin zone given by
\begin{equation}
\fl\Delta \Smr = \frac{1}{2Ns} \smk \sum_\alpha \lf\langle \cdak{}\cak{} + 
                 \cdak{-}\cak{-} \rt\rangle
               = \frac{1}{2Ns} \smk \lf\langle \Nd{X}\N{X}\rt\rangle -\frac{1}{2}.
\label{DeltaS}
\end{equation}
Recall that we have introduced new boson operators $\ak{}$ which diagonalize the
spin wave Hamiltonian through the canonical Bogoliubov transformation 
$\N{X} = \N{Q}\N{Y}$, where it is convenient to note that the transformation can 
be written in component form as $\N{X}_\alpha = \sum_i \Qc{\alpha}{i} \N{Y}_i$ and 
$\N{X}_{\alpha+s}  =  \sum_i  \Qc{\alpha+s}{i}\N{Y}_i$ 
with the convention that Greek indices run between $1 \ldots s$ and Latin indices between
$1 \ldots 2s$.  The inner product, $\Nd{X}\N{X}$ can then be expanded as
\begin{eqnarray}
\fl\Nd{X}\N{X} = 
      \sum_i \smab \lf[ \Qdc{\alpha}{i}\Qc{i}{\beta}\adks{}{\alpha}\aks{}{\beta} +
      \Qdc{\alpha+s}{i}\Qc{i}{\beta+s}\adks{-}{\beta}\aks{-}{\alpha} \rt. \nonumber\\
       \lf. \lo{+} \Qdc{\alpha}{i}\Qc{i}{\beta+s}\adks{}{\alpha}\adks{-}{\beta} +
      \Qdc{\alpha+s}{i}\Qc{i}{\beta}\aks{-}{\alpha}\aks{}{\beta} + 
      \Qdc{\alpha+s}{i}\Qc{i}{\beta+s}\dlta{\alpha}{\beta} \rt] 
\label{sumca-exp}
\end{eqnarray}      
and at zero temperature there are no thermally activated spin deviations to destroy with $\ak{}$, 
so the expectation value of all quadratic boson terms tend to zero.  Substituting
\Eref{sumca-exp} into (\ref{DeltaS}) and neglecting these terms we find the 
staggered moment at zero temperature due to quantum spin fluctuations to be
\begin{equation}
\Delta \Smr = \frac{1}{2}\lf( \frac{1}{Ns} \smk \sum_\alpha \lf[\Nd{Q}\N{Q}\rt]_{\alpha \alpha}
		  - 1 \rt).
\label{gs-cor-moment}
\end{equation}

At this point we may reflect upon the fact that beginning with \Eref{H-mat} the form
of the fully diagonalized spin wave Hamiltonian to $\Or(1/\Smr)$ in \Eref{QBSWH} could have been 
inferred with little or no calculation. Physically, to order 1/S, the Hamiltonian ought to
describe a system of non-interacting bosons with energy dispersion equal to 
$\hbar\omega_{\alpha}(\vk)$ where $\omega_{\alpha}(\vk)$ 
are the classical spin wave frequencies obtained by linearizing the classical equations 
of motion for interacting magnetic dipoles.  Self contained approaches exist which begin
from such a Hamiltonian and allow for the calculation of the staggered magnetic moment
at zero temperature without any knowledge of the eigenvectors of the diagonalizing 
transformation embedded within the $\N{Q}$ matrix discussed above.  A staggered 
anisotropy field $h_{\alpha}$ is introduced which points 
along the local quantization (classical ground state) direction and the change in the 
ground state energy $\partial \mathrm{E} / \partial h_{\alpha}$ is calculated giving 
the total magnetic moment on each sublattice at $T=0$ K \cite{corr-white}.  
In the discussion above, we have chosen not to follow this physically motivated approach
and have instead utilized the full spectrum of our transformation matrix $\N{Q}$
to compute the desired zero-point quantum effects.


\subsection{Thermodynamic Relations}
We have decomposed our theory of spins on a lattice interacting via frustrated exchange 
and dipole-dipole interactions to that of non-interacting Bose gas (\Eref{QBSWH}). 
Beginning from the well known partition function,
\begin{equation}
\mc{Z} = \Tr\exp[-\beta\mc{H}] = 
       \exp[-\beta\mathrm{E}]\prod_{\bi{k}}\prod_{\alpha}
	    \frac{1}{1-\exp[-\beta\ek{}]}
\label{Z}  
\end{equation}
where $\mathrm{E} =  \mc{H}^{(0)} + \Delta\mc{H}^{(0)}$, and $\beta$ is the inverse
temperature we have used the standard thermodynamic relations to derive equations 
for the free energy ($\mc{F}$), average internal energy ($\mc{U}$), specific heat 
at constant volume ($\mc{C}_v$) and the entropy ($\mc{S}$) as a function of temperature. 
These expressions will be valid only for low temperatures where $T \ll \ek{}$ and are given by 
\begin{equation}
\fl\eqalign{
\mc{F} = \rm{E} + \beta^{-1}\smk\sma\ln\lf(1-\exp[\beta\ek{}] \rt) \qquad &
\mc{U} = \rm{E} + \smk\sma \ek{}n_{\rm{B}}(\ek{}) \\
\mc{C}_v = \beta^2\smk\sma[\ek{}n_{\rm{B}}(\ek{})]^2\exp[\beta\ek{}] \qquad &
\mc{S} = \beta(\mc{U}-\mc{F})}
\label{thermos}
\end{equation}
where $n_{\rm{B}}(\ek{}) = \lf(\exp[\beta\ek{}]-1\rt)^{-1}$ is the Bose factor
\cite{pathria}.
In order to calculate the magnetic order parameter as a function of 
temperature ($m^{\star}$), we must return to the general form for the reduction of the sublattice 
magnetization in \Eref{sumca-exp}. For suitably low temperatures, the expectation value 
of all non-number conserving quadratic boson terms will be negligible and we find
\begin{equation}
m^{\star} = \Smr-\Delta\Smr(T=0) -
\frac{1}{Ns} \smk\sma \left[\Nd{Q}\N{Q}\rt]_{\alpha\alpha}n_{\rm{B}}(\ek{}).
\label{mstar}
\end{equation}

\section{The Heisenberg-Like Dipolar Pyrochlore}
\label{HLP}
Having developed the formalism of linear spin wave theory for a multi-sublattice
non-Bravais magnetic system we now map out the low temperature excitations of spins 
interacting via nearest neighbour Heisenberg exchange and long range dipole-dipole 
interactions on the pyrochlore lattice.  

\subsection{Classical Ground States}

It is known that when isotropic nearest neighbour exchange interactions alone are
considered on the pyrochlore lattice, the system exhibits
no global symmetry breaking and has soft modes at long wavelengths
\cite{Can-Lac,Moes-Chalk}. The inclusion of a perturbation such as long-range 
dipole-dipole interactions breaks the continuous O(3) symmetry \cite{gdtio-huse,Palm-Chalk}
and leads to a gap in the excitation spectrum.  The non-trivial coupling of 
the isotropic spin symmetry and translational lattice symmetry for a particular set 
of interaction parameters controls how these optical modes disperse.  The details of 
this dispersion directly affect the stiffness of a given spin configuration and 
quantitatively determines the level of proliferation of quantum fluctuations in the ground state.

\begin{figure}[ht]
\begin{center}
\includegraphics[scale=0.75]{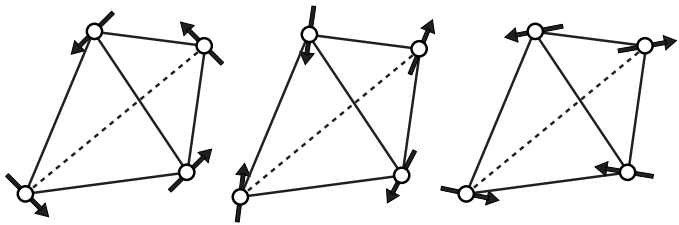}
\end{center}
\caption{\label{PC} Six degenerate Palmer-Chalker (PC) ground states 
(reverse spins for other three) for spins on a single tetrahedron.}
\end{figure}

Palmer and Chalker have investigated the classical Heisenberg model for spins residing 
on the pyrochlore lattice with long-range dipolar interactions \cite{Palm-Chalk}.  
They find that the dipolar perturbation fixes the two continuous internal degrees of freedom and 
leads to the energetic selection of the six degenerate $\bi{q}=0$ (Palmer-Chalker, PC) 
ground states shown in Figure~\ref{PC} where the total magnetic moment on each tetrahedron is zero.  
We have confirmed the six degenerate PC classical ground states numerically 
in real space using a field averaging approach by exploring the energy landscape 
at zero temperature as shown in Table~\ref{pyro-sgs}.  The inclusion of long range 
dipole interactions does indeed break the global rotational symmetry and selects a 
crystallographic plane in which the spins are confined while still preserving a zero 
moment structure on each tetrahedron.  
\begin{table}[ht]
\caption{\label{pyro-sgs} Vectors which define the six classical PC ground state spin
         configurations on the pyrochlore lattice.}
\begin{indented}
\item[]\begin{tabular}{@{}lll}
\br
PC$_{xy}$ State & PC$_{xz}$ State & PC$_{yz}$ State \\
\mr
$\bi{S}_1$ = $1/\sqrt{2} (1,\overline{1},0)$ &
$\bi{S}_1$ = $1/\sqrt{2} (1,0,\overline{1})$ &
$\bi{S}_1$ = $1/\sqrt{2} (0,1,\overline{1})$ \\

$\bi{S}_2$ = $1/\sqrt{2} (\overline{1},1,0)$ & $\bi{S}_2$ = $1/\sqrt{2} (1,0,1)$ &
$\bi{S}_2$ = $1/\sqrt{2} (0,1,1)$ \\

$\bi{S}_3$ = $1/\sqrt{2} (1,1,0) $ &
$\bi{S}_3$ = $1/\sqrt{2} (\overline{1},0,1) $ &
$\bi{S}_3$ = $1/\sqrt{2} (0,\overline{1},\overline{1}) $ \\

$\bi{S}_4$ = $1/\sqrt{2} (\overline{1},\overline{1},0)$ &
$\bi{S}_4$ = $1/\sqrt{2} (\overline{1},0,\overline{1})$ &
$\bi{S}_4$ = $1/\sqrt{2} (0,\overline{1},1)$ \\
\br
\end{tabular}
\end{indented}
\end{table}
The discrete degeneracy of the PC ground states allows us to perform our
spin wave calculation on only one state, and the remainder of the results presented were
calculated for the PC$_{xy}$ state, but are identical (with the proper symmetry 
considerations) for the other five states.

\subsection{Spin Wave Dispersion}

The functional dependence of the normal modes of an oscillating system on the wavevector
can be derived from classical mechanics. Treating each spin as a classical rotor, 
the rate of change of the angular momentum in the local spin frame, 
$\hbar\partial\Svrat / \partial t$ is equal to the torque,
$\tilde{\bi{N}}_\alpha (\vu{R}{\mu}) = \tilde{\bi{\mu}}_\alpha (\vu{R}{\mu}) \times 
\tilde{\mathrm{\bi{B}}}_\alpha (\vu{R}{\mu})$ acting on $\Svrat$, where
$\tilde{\mathrm{\bi{B}}}_\alpha (\vu{R}{\mu})$ is the total effective magnetic field
experienced by a spin on site $\alpha$ in the local frame.  Here we must take care 
to ensure that the vector product is carried out in a way that is independent of the local 
coordinate system.  Following this procedure we use a plane-wave ansatz
of the form $\Scra{j} \sim \exp{[i\vec{k}\cdot(\vu{R}{\mu}+\ra) - i\omega t]}$ to solve the 
resulting system of linear equations and obtain the excitation spectrum.  The spin wave
dispersion spectrum calculated using Holstein-Primakoff bosons above can then be compared 
with the classical modes of excitation as an essential consistency check.

Armed with the knowledge of a set of candidate classical ground state spin
configurations for the isotropic dipolar pyrochlore (which satisfy $\mc{H}^{(1)} = 0$)
we may now attempt to apply this method to an actual material, the frustrated pyrochlore 
antiferromagnet gadolinium titanate.  $\Gd$ is a spin only ($\Smr = 7/2$) material with 
an isotropic magnetic exchange parameter between nearest neighbours which can 
be estimated from the value of the Curie-Weiss constant ($\theta_{\rm CW} = -9.8$~K) using the relation 
$J_1 = 3\theta_{\rm CW}/[z\Smr(\Smr+1)]$ $\simeq -0.30476\ \mbox{K}$ 
where $z$ is the number of nearest neighbours.  The dipolar energy scale is set at 
$\Ddd / (a\sqrt{2}/4)^3 \simeq 0.05338$~K where $a=10.184\ \mbox{\AA}$ 
is the edge size of the cubic unit cell and $|\boldsymbol{\mc{R}^{1}}| = a\sqrt{2}/4$ is the 
nearest neighbour distance leading to $(\Ddd|\boldsymbol{\mc{R}^{1}}|^{-3}) / |J_1| \sim 1/5$ 
\cite{gdtio-raju}. 

In order to map out the low temperature magnetic excitations of this material we are 
faced with the non-trivial task of calculating the Fourier transform of the $12\times12$
interaction matrix $\J{\R}$ (\Eref{H-2})  which contains the exchange contribution 
out to a finite number of nearest neighbours, and the infinite range dipole-dipole
interactions. This is accomplished via the Ewald summation method \cite{ewald}
and the final result for up to third nearest neighbour isotropic exchange
\cite{agrd-thesis,ewald-calc} is given by
\begin{equation}
\J{\bi{k}} = \E{\bi{k}} - \Ddd\D{\bi{k}}
\end{equation}
where
\begin{eqnarray}
\fl\E{\bi{k}} = 2J_1(\nhia\cdot\nhjb)\cos[\bi{k}\cdot(\ra-\rb)](1-\dlta{\alpha}{\beta}) \nonumber \\
   \lo{+} 2J_2 (\nhia\cdot\nhjb) \sum_{\gamma\ne\alpha,\beta}\cos[\bi{k}\cdot(\ra+\rb-2\vd{r}{\gamma})]
          (1-\dlta{\alpha}{\beta}) \nonumber \\
      \lo{+} 2J_3 (\nhia\cdot\nhjb) \sum_{\sigma<\rho}\cos[ 2 \bi{k}\cdot(\vd{r}{\sigma}-\vd{r}{\rho})]
              \dlta{\alpha}{\beta}
\end{eqnarray}
and
\begin{eqnarray}
\fl\D{\bi{k}} = -\frac{4\lambda^3}{3\sqrt{\pi}}(\nhia\cdot\nhjb)\dlta{\alpha}{\beta} +
	       \frac{4\pi}{v}\sum_{\bi{G}}\K{\bi{k}-\bi{G}}
	       \mathrm{e}^{-\nrm{\bi{k}-\bi{G}}{2}\lf/4\lambda^2\rt.}
	       \edot{-}{\bi{G}}{\rab} \nonumber \\
	       \lo{+} \sum_{\mu}{}'\lf[\U{1}{\R}-\U{2}{\R}\rt]\edot{-}{\bi{k}}{\R}
\end{eqnarray}
with
\numparts
\begin{eqnarray}
\fl\K{\bi{k}-\bi{G}} &= \frac{\lf[\nhia\cdot(\bi{k}-\bi{G})\rt]\lf[\nhjb\cdot(\bi{k}-\bi{G})\rt]}
                        {\nrm{\bi{k}-\bi{G}}{2}} \\
\fl\U{1}{\R} &= \nndot\frac{1}{\nrm{\R}{2}}\lf[ \frac{2\lambda}{\sqrt{\pi}}\mathrm{e}^{-\lambda^2\nrm{\R}{2}}
	      + \frac{\mathrm{erfc}(\lambda\nrm{\R}{})}{\nrm{\R}{}} \rt] \\
\fl\U{2}{\R} &= \nrdot \frac{1}{\nrm{\R}{2}} \nonumber \\
             &\quad\times \lf\{ \frac{2\lambda}{\sqrt{\pi}} \lf[ 2\lambda^2 +
                     \frac{3}{\nrm{\R}{2}} \rt] \mathrm{e}^{-\lambda^2\nrm{\R}{2}} 
	            + 3\frac{\mathrm{erfc}(\lambda\nrm{\R}{})}{\nrm{\R}{3}} \rt\}.
\end{eqnarray}
\endnumparts
In these expressions $\lambda=\sqrt{\pi/v}$ is the Ewald convergence parameter,
erfc($\cdots$) is the complimentary error function,  $v$ is the volume of the primitive unit cell, 
$\vg$ is a FCC reciprocal lattice vector and the primed summation indicates that the 
real-space sum is over all FCC Bravais lattice vectors \emph{except} $\R=0$.

\begin{figure}[ht]
\begin{center}
\includegraphics[scale = 0.7]{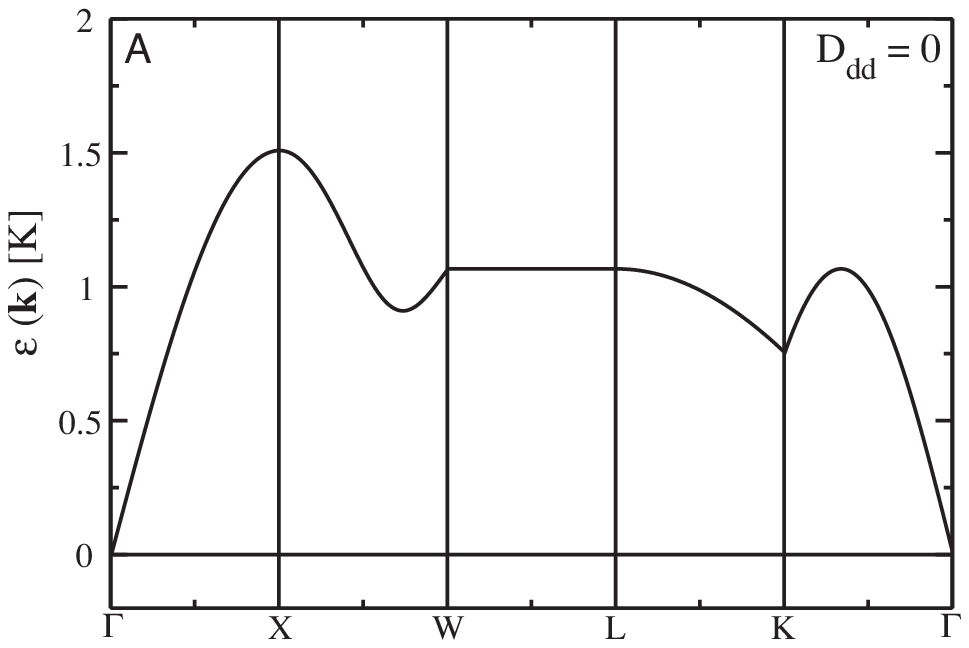}
\includegraphics[scale = 0.7]{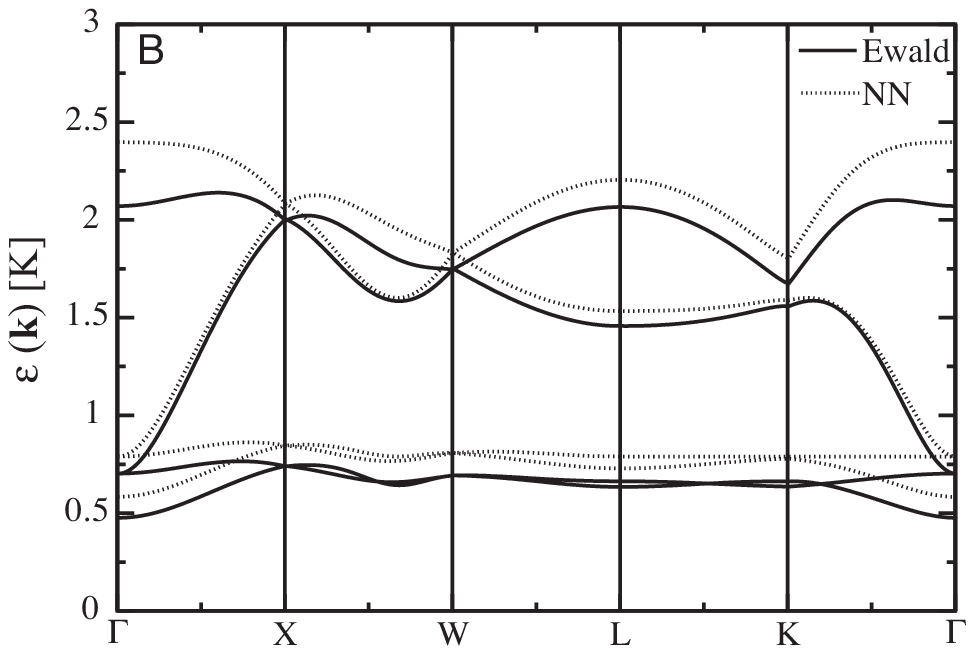}
\end{center}
\caption{\label{spec} Spin-wave energies, $\varepsilon_{\alpha}(\bi{k})$ in kelvin, for 
(A) the pure Heisenberg pyrochlore ($\Ddd = 0$) with $J_1 = -0.30476$~K where both branches
are doubly degenerate and  (B) excitations from one of six discretely degenerate 
classical ground states for $\Gd$ with nearest neighbour exchange and both nearest neighbour 
and long range dipole interactions using the Ewald method.}
\end{figure}

If we completely ignore the effect of dipole interactions we expect the excitation
spectrum of a system with massless Goldstone bosons as seen in Figure~\ref{spec} (A). 
There are two unfixed internal degrees of freedom associated with the spin 
isotropy and one time reversal degree of freedom.  The ground state spin configuration can 
fluctuate with no cost in energy between states in its continuously degenerate manifold 
while maintaining a macroscopic zero-moment structure.  

When the effects of dipole-dipole interactions are included, the internal degrees of
freedom are fixed, and the spins are able to minimize their energy by ordering in a
non-collinear, planar configuration which satisfies a pair-wise antiferromagnet constraint.
This can be done in a number of ways, provided that each spin is parallel to an opposite
edge of the tetrahedron (refer to Figure~\ref{PC}) and we move from a continuously
degenerate manifold to one with the six Palmer-Chalker discrete ground states.

Beginning from the $PC_{xy}$ state with isotropic nearest neighbour exchange ($J_2 = J_3 = 0$) 
the spin wave dispersion spectrum for the parameter set of $\Gd$ is shown in 
Figure~\ref{spec} (B).  We observe four optical modes indicating that all distortions from 
the chosen ground state are associated with a positive energy cost. As predicted, the 
introduction of anisotropy has broken a global symmetry and lifted the spectrum to finite frequency.  
The remnant soft modes from the pure Heisenberg model (Figure~\ref{spec} (A)) have remained 
moderately degenerate and relatively dispersionless, while the acoustic modes have split and 
are dispersing to the point where one mode has changed curvature along $\Gamma \to X$
(for the $PC_{xy}$ state).  Once the system is in one of its stable classical ground states, 
it is unable to move between equivalent energy states through a low energy 
long-wavelength acoustic mode. This implies that we were justified in making our
semi-classical approximations as it is likely the $\Gd$ may display very little quantum
mechanical behaviour to asymptotically low temperatures.

The inclusion of truly long range dipole interactions using the Ewald technique, as 
opposed to truncating at nearest neighbour, leads to a restoration of some 
previously buried lattice symmetry.  In Figure~\ref{spec} (B) we observe promotion of 
quasi-degeneracy in the excitation spectrum between $W$ and $K$, and a slight suppression of 
the anisotropy gap throughout the Brillouin zone due to the global dipolar field being 
partially screened.  It is interesting to note that when interaction terms beyond nearest 
neighbour are neglected, quantum fluctuations are inhibited in the ground state as 
the local unscreened interaction leads to an increased anisotropy gap at the zone 
center; $\varepsilon(\vk)$ using the Ewald method is lower than 
$\varepsilon(\vk)$ using only nearest neighbour interactions at $\Gamma$.

\subsection{Quantum Spin Fluctuations}

To form a quantitative picture of the role of spin fluctuations in the ground state 
we must calculate both the shift of the classical ground state energy and the reduction of
the fully polarized $\Smr=7/2$ moment in $\Gd$ at zero temperature.
This is accomplished using the properly normalized eigenvectors of the numerically
constructed Bogoliubov transformation in conjunction with Equations~(\ref{H-0}) and 
(\ref{DeltaH}) for the energy of the system and ~\Eref{gs-cor-moment} for the 
reduced moment. Performing averages over the first Brillouin zone of the FCC lattice, 
the main quantitative numerical results of this study, accurate to order $1/\Smr$ are 
\numparts
\begin{eqnarray}
\label{qcora}
\rm{E}_{NN} = \rm{E}_0 (1+0.1442/\Smr) \qquad & m^\star_{NN} = \Smr(1-0.1007/\Smr) \\
\label{qcorb}
\rm{E}_{LR} = \rm{E}_0 (1+0.3526/\Smr) \qquad & m^\star_{LR} = \Smr(1-0.1174/\Smr)
\end{eqnarray}
\endnumparts
where $\rm{E}_0 = -1/2N\Smr^2\smab\Jc{z}{z}{0}$ is the classical ground state
energy, $NN\equiv\mbox{Nearest Neighbour Dipole}$ and $LR\equiv\mbox{Long Range Dipole}$.  In accordance 
with our naive expectation from an examination of the spin wave dispersion spectrum, 
we see a softening of the system and enhanced spin fluctuations in the ground state when
long range dipole interactions are treated properly.  The reduction of the fully 
polarized moment is almost twenty percent larger when the Ewald summation method is
employed.  These comparisons are overshadowed by the reality that the
actual numerical values of the quantum corrections are negligible when the zero
point shifts above are divided by the large $\Smr=7/2$ moment of the Gd$^{3+}$ ion.  The
ground state energy is shifted by less than ten percent and the fully ordered moment is
reduced by half of that.  However, the $\Smr$-dependence of the numerical values for the 
quantum corrections in \Eref{qcora} and (\ref{qcorb}) has been explicitly removed, and 
these numbers are set by $J$ and $(\mu_{0}/4\pi)(\gub)^2/\nrm{\boldsymbol{\mc{R}^1}}{3}$ for a 
given interacting system.  It is apparent that one can not a priori rule out the possibility of 
sizeable quantum fluctuations ($\sim 5\%$ for $\Smr=7/2$ here) in the ground state of the frustrated 
pyrochlore materials.  Yb$_2$Ti$_2$O$_7$ falls into the category of an XY $\Smr = 1/2$ system 
\cite{YbTiO} and using \Eref{qcora} and (\ref{qcorb}) above we might expect quantum fluctuations on 
the order of 20\%.  Therefore, it is crucial to properly identify and include the variety of complex
competing interactions which act as finite, non-zero perturbations to the isotropic Heisenberg 
exchange parameter $J$.

\subsection{Thermodynamic Properties}
\begin{figure}[ht]
\centering
\includegraphics[scale=0.6]{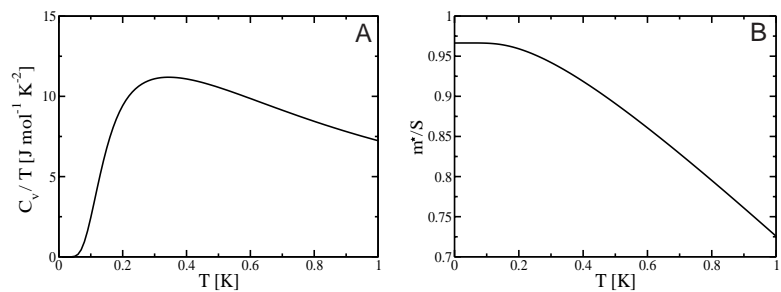}
\caption{\label{thermo} The specific heat at constant volume scaled by the temperature (A) 
and the staggered magnetic order parameter (B) as a function of temperature calculated using 
linear spin wave theory for gadolinium titanate.} 
\end{figure}

We have calculated the specific heat and staggered magnetization at finite temperatures and
the results are shown in Figure~\ref{thermo}.
At low temperatures, all four spin wave modes are gapped and we observe
an exponentially vanishing specific heat as $T \rightarrow 0$~K unlike the 
$T^2$ laws observed for both $\Gd$ \cite{gdtio-huse} and Gd$_2$Sn$_2$0$_7$ \cite{GdSnO}.  
It appears that the energy scale set by the anisotropy gap corresponds to the peak 
position near 0.4 K in panel (A) and there exists some qualitative similarities 
with the experimentally measured value of $\mathcal{C}_v/T$ \cite{gdtio-huse} where 
two peaks are observed at approximately 0.7~K and 1.0~K. However, we have confined the system 
to a particular classical ground state and cannot probe the existence of multiple phase 
transitions in this study \cite{gdtio-huse,gdtio-bram,gdtio-matt,gdtio-shas}. 

The temperature dependence of the staggered magnetization shown in Figure~\ref{thermo} panel 
(B) is also of interest and we find the behaviour, $1-\exp[-\Delta/T]T^3$, with
$\Delta \sim \mbox{gap}$, as expected for an antiferromagnet at low temperature
\cite{magnetism-I}.

\section{Conclusions}
We began by examining the energy landscape at zero temperature
for the dipolar Heisenberg pyrochlore system and have confirmed the existence of the
six-fold degenerate Palmer-Chalker ground states. Beginning from one of these classical
ground states we have employed the techniques of non-interacting spin wave theory.
Using bosonic spin-deviation operators and an analysis of the quantum
mechanical equations of motion we have numerically constructed the canonical
Bogoliubov transformation and diagonalized the spin wave Hamiltonian for the interaction
parameter set of $\Gd$. We have obtained the spin wave dispersion spectrum for nearest 
neighbour exchange and both nearest neighbour and long range dipole interactions using the 
Ewald summation technique and confirmed it classically treating each spin as a rotor in 
its local frame. The spin wave excitation spectrum displays four non-degenerate modes 
and is gapped throughout the first zone.  Ensuring that a proper bosonic 
transformation matrix has been found for each wavevector $\bi{k}$, we have 
calculated the numerical contribution of zero point quantum spin fluctuations to the 
ground state energy and the reduction of the classical staggered moment by performing a 
discrete average over the first Brillouin zone.  The partition function was calculated using
the fully diagonalized spin wave Hamiltonian and we obtained equations for the most relevant 
thermodynamic relations at low temperatures.

We have confirmed the stability of the PC ground states against quantum fluctuations and
it appears that due to the significant anisotropy gap present throughout the zone and 
the large value of the classical spin $\Smr = 7/2$ of the gadolinium ion, quantum 
mechanical effects are small ($\sim$10\%) in this system. The fully ordered spin 
polarization is reduced by less than five percent of its classical value, and it should be
well described by classical methods at low temperature. The physical system retains little 
`quantum mechanical' knowledge of its fully disordered paramagnetic state after the 
effects of dipole interactions have been included in the Hamiltonian. 

Small perturbations arising from exchange interactions beyond nearest neighbour 
may also be significant in this system, and recent mean field theory studies of $\Gd$ with 
$J_2 \ne J_3 \ne 0$ have identified another possible kagom\'{e}-like ground state with ordering wavevector
$\bi{q}=\lf(\frac{1}{2}\frac{1}{2}\frac{1}{2}\rt)$.  This state has one sublattice with a net moment
of zero and it appears to be energetically selected over the PC states above some crossing 
temperature $T_{cross}$ \cite{gdtio-matt}.  Powder neutron scattering results for $\Gd$
\cite{gdtio-bram} and recent theoretical work by C\'{e}pas and Shastry
\cite{gdtio-shas} would tend to confirm the existence of this $1/4$-disordered
ground state at temperatures below 1 K. In contrast, an experimental study of the frustrated pyrochlore 
material Gd$_2$Sn$_2$O$_7$ \cite{GdSnO} shows a single sharp first order transition to a long-range 
ordered magnetic state near 1~K which may be PC in character. It is therefore still under
debate whether the real low temperature ordered phase of these Heisenberg-like dipolar pyrochlore 
materials is well described by the PC manifold. Nevertheless, the stability of the six-fold degenerate
classical PC states against quantum fluctuations in these systems is of considerable interest.

Although we have chosen to limit our calculation to one specific
material, gadolinium titanate, we have presented the calculation in a general form with no
real dependence on the structure of the lattice, or the spin symmetry. 
It is therefore hoped that the theoretical techniques presented in this study can be 
employed to provide further understanding of the peculiar and 
interesting properties of highly frustrated antiferromagnets on non-Bravais lattices.

\ack
The authors would like to thank Steve Bramwell, Jean-Yves Delannoy, Matthew Enjalran, 
Ying-Jer Kao, Roger Melko, and Hamid Molavian for many useful discussions. This work
is supported by NSERC of Canada, Research Corporation, the Province of Ontario and the
Canada Research Chair Program.

\appendix
\section{Numerically Constructed Bogoliubov Transformation}
\label{ncbt}
The equations governing the construction of $\N{Q}$ can be derived \cite{white-diag}
by utilizing the vector commutator
\begin{equation}
\lf[\N{X},\Nd{X}\rt] \equiv \N{X}\tilde{\N{X}}^{T} - \lf(\tilde{\N{X}}\N{X}^{T}\rt)^{T},
\end{equation}
where $\tilde{\N{X}}$ is the column vector of adjoint boson operators and $T$ indicates
the usual transpose operation. Using this definition we find
\begin{equation}
\lf[\N{X},\Nd{X}\rt]  = \N{g}.
\label{XXdg}
\end{equation}
where the metric of the transformation is
\begin{equation}
\N{g} = 
\lf[\begin{array}{cc}
\mathbb{1} & 0 \\
0          & -\mathbb{1} \\
\end{array}\rt]
\end{equation}
and $\mathbb{1}$ is the $s \times s$ identity matrix where $s$ is the 
number of magnetic sublattices.  The required ortho-normalization condition for 
the transformation matrix $\N{Q}$ is found to be
\begin{equation}
\N{Q} \N{g} \Nd{Q} = \N{g}.
\label{QgQdg}
\end{equation}
The hermitian conjugate of $\N{Q}$ can be expressed in terms of its inverse as
$\Nd{Q} = \N{g} \N{Q}^{-1} \N{g}$ and the diagonalization condition is now given by
$\Nd{Q} \N{H} \N{Q} = \N{g} \N{Q}^{-1} \N{g} = \bLambda$
leading to the more familiar eigenproblem
\begin{equation}
\N{Q}^{-1} \N{g} \N{H} \N{Q} = \N{g}\bLambda.
\end{equation}
Defining $\N{L} = \N{g}\N{H}$ and $\bLambda_\mathsf{L} = \N{g}\bLambda$ we
can use a standard numerical package to solve for the eigenvectors and eigenvalues 
of the non-hermitian matrix $\N{L}$, allowing us to write
$\N{Q_L} \bLambda_\mathsf{L} \N{Q_L}^{-1} = \N{L}$
where $\N{Q_L}$ is the matrix whose columns are the eigenvectors of $\N{L}$ subject to
normalization with respect to the usual identity metric.

In order to find the proper diagonalizing transformation $\N{Q}$ at each $\vk$-vector in the
first Brillouin zone, three conditions must be met:
\begin{displaymath}
\begin{array}{cccc}
 \mathrm{(I)}   & \N{Q} \bLambda_\mathsf{L} \N{Q}^{-1} & = &  \N{L} \\
 \mathrm{(II)}  & \N{Q} \N{g} \Nd{Q} & = &  \N{g} \\
 \mathrm{(III)} & \Nd{Q} \N{H} \N{Q} & = &  \bLambda.
\end{array}
\end{displaymath}
These can be satisfied simultaneously, provided that we construct a Bogoliubov
transformation which imposes the full \emph{bosonic constraint} on our transformed operators.
This amounts to properly grouping, organizing and then renormalizing the computed
eigenvectors of $\N{L}$ through a block transformation, ensuring that we always remain
within only one eigenspace at a time \cite{jy-diluted}.

Suppose $\N{L}$ has $m$ unique eigenvalues, each having degeneracy $d_j$ for
$j=1,\ldots,m$. We assume that our eigenvector matrix $\N{Q_L}$ is organized such that
eigenvectors belonging to the same eigenspace are grouped together.  Our goal is to construct
a block diagonal transformation matrix $\N{dB}$ consisting of $m\; d_j \times d_j$ blocks which will
transform $\N{Q_L}$ to our desired matrix $\N{Q}$ via $\N{Q} = \N{Q_L} \N{dB}$.
By performing this transformation we have simply formed new linear combinations 
of the eigenvectors within a given eigenspace, and thus condition (I) will be satisfied. 
Condition (II) requires that
\begin{displaymath}
\N{(Q_L dB)}\N{g}\Nd{(Q_L dB)} = \N{g}
\end{displaymath}
and provided that there are only gapped spin wave excitations (no zero eigenvalues), $\N{Q_L}$ 
is non-singular so we can rearrange this as
\begin{equation}
\N{dB}\N{g}\Nd{dB} = \N{M},
\end{equation}
where we have defined the block-diagonal matrix $\N{M}=\lf(\Nd{Q_L}\N{g}\N{Q_L}\rt)^{-1}$.
For each block $\N{dB}$ we must solve an equation of the form
\begin{equation}
\N{g}_j \N{dB}_j \Nd{dB}_j = \N{M}_j
\label{gdBdBdM}
\end{equation}
where
\begin{equation}
\N{g}_j=\cases{ 1& if $\sum_{i<j} d_j < s$ \\
                -1 & otherwise \\},
\end{equation}
\begin{equation}
\N{M}_j = \lf(\Nd{Q}_{\mathsf{L},j}\ \N{g}\ \N{Q}_{\mathsf{L},j} \rt)^{-1}
\end{equation}
and $\N{Q}_{\mathsf{L},j}$ is the $2s \times d_j$ block of vectors belonging to eigenspace
$j$.  An equation of of this type can be solved by decomposing $\N{M}_j = \N{V}_j \N{D}_j
\N{V}_j^{-1}$
where $\N{V}_j$ is the unitary eigenvector matrix and $\N{D}_j$ is the diagonal eigenvalue
matrix corresponding to $\N{M}_j$.  Substituting this into \Eref{gdBdBdM} we find that 
\begin{equation}
\N{dB}_j = \N{V}_j \sqrt{\N{g}_j\ \N{D}_j}\ \N{V}_j^{-1}
\end{equation}
is always a solution.  Condition (III) is finally satisfied and we have arrived at 
an equation which can be used to numerically construct each block of $\N{dB}$ 
yielding the proper transformation matrix $\N{Q}$ for each $\vk$.


\end{document}